# 3D non-LTE line formation in the solar photosphere and the solar oxygen abundance

Dan Kiselman[1,2] and Åke Nordlund[3,4]

[1] NORDITA, Blegdamsvej 17, DK-2100 Copenhagen Ø, Denmark
[2] Royal Swedish Academy of Sciences, Stockholm Observatory, S-133 36 Saltsjöbaden, Sweden
[3] Theoretical Astrophysics Center, Blegdamsvej 21, DK-2100 Copenhagen Ø, Denmark
[4] Astronomical Observatory, Niels Bohr Institute for Astronomy, Physics and Geophysics, Øster Voldgade 3, DK-1350 Copenhagen K, Denmark

**Abstract.** We study the formation of O I and OH spectral lines in three-dimensional hydrodynamic models of the solar photosphere. The line source function of the O I 777 nm triplet is allowed to depart from local thermodynamic equilibrium (LTE), within the two-level-atom approximation. Comparison with results from 1D models show that the 3D models alleviate, but do not remove, the discrepancy between the oxygen abundances reported from non-LTE work on the 777 nm triplet and from the [O I] 630 nm and OH lines. Results for the latter two could imply that the solar oxygen abundance is below 8.8. If this is confirmed, the discrepancy between theory and observation for the 777 nm triplet lines might fall within the range of errors in equivalent-width measurements and $f$-values. The line source function of the 777 nm triplet in the 1.5D approximation is shown to differ insignificantly from the full 3D non-LTE result.

## 1. Introduction

### 1.1. Background

The commonly used theoretical and semi-empirical models of the solar photosphere are one-dimensional, implying homogeneous and plane-parallel stratification. The real solar photosphere does not look like this. The phenomenon of solar granulation shows us an evolving pattern of bright rising gas in granules and falling gas in darker intergranular lanes. Numerical hydrodynamic simulations of granulation provide both qualitative insight and quantitative information about a solar photosphere very far from thermal homogeneity. A one-dimensional model represents an averaged atmospheric structure, and it seems unlikely that one single such model can suffice to represent all aspects of the photosphere. An obvious example is the asymmetries of Fraunhofer lines that cannot be reproduced using one-dimensional photospheric models. Three-dimensional hydrodynamic granulation simulations have proven successful in reproducing the shapes of different lines (see, e.g., Dravins et al. 1981 and Dravins & Nordlund 1990a, 1990b).

The question of the importance of granulation for the strength of the lines in integrated light has not been addressed much. It is not clear whether the shortcomings of one-dimensional photospheric models can seriously challenge, for example, derivations of elemental abundance ratios.

Holweger et al. (1990) argued, on the basis of simulation results by Steffen (1989), that the varying temperature gradient in the granulation pattern should be more important for the strength of photospheric lines than the temperature variations themselves. Thus all lines should behave qualitatively in the same way and the line ratios, from which we determine abundance ratios, should be essentially unchanged over the granulation pattern. Kiselman (1994) verified this observationally for a number of lines from different elements. This result should, however, be regarded as absence of evidence for the existence of very large systematic errors in abundance determinations, rather than evidence for their absence. The problem of quantifying the errors remains, as does the possibility that some lines will indeed behave very differently. One could, for example, suspect that lines which are predicted to be formed far from local thermodynamic equilibrium, LTE, are sensitive to granulation.

The fact that spectral analysis of stars seems to give consistent results should not be taken as proof that er-dard analysis employs several free parameters – notably enhancement factors of pressure damping and Gaussian microturbulence (also macroturbulence when line profiles are studied). These are adjusted until derived abundances, etc., are consistent, but it is possible that the discrepancies one tries to get rid of are inherent to the model approximations of LTE and homogeneity. Microturbulence may not have much to do with physical velocity fields in the atmosphere, and enhancement factors in the pressure damping may not really reflect shortcomings in atomic line-broadening theory. Examples of this are supersonic microturbulent velocities for hot stars deemed unnecessary by non-LTE spectral line modelling (Becker & Butler 1988, 1989), and the observation of Nordlund (1984) that the velocity fields in the solar photosphere may cause line wings that need extra damping if one assumes the velocity fields (micro- and macro-turbulence) to be Gaussian.

*1.2. The oxygen problem*

Oxygen is a cosmically abundant and astrophysically important element. Efforts to measure oxygen abundances in solar-type stars of different metallicities in order to trace the chemical evolution of the Galaxy have led to somewhat discrepant results depending on the lines observed. LTE analyses seem to give systematic differences between the abundances derived for the same stars using the [O I] line at 630 nm and the O I triplet at 777 nm (Magain 1988, Spite & Spite 1991). The triplet gives higher abundances and Nissen & Edvardsson (1992) found the difference to have a tendency to grow with decreasing metallicity. There may be sufficient freedom within the usual parameters in stellar spectral analysis to reconcile the differences within the approximation of LTE and plane-parallel atmospheres, for example by adjusting the effective temperature scale (King 1993), but there is a discrepancy also for the Sun where the freedom of parameters is more restricted.

The solar discrepancy starts with the observation that the centre-to-limb variation of the 777 nm triplet is incompatible with the lines being formed in LTE (Altrock 1968), while non-LTE modelling of these lines reproduces the disk behaviour very well (Kiselman 1991, 1993). However, non-LTE work that reproduces the disk variation also predicts line strengths that are too strong for an oxygen abundance around 8.9[1] that is derived from the [O I] 630 nm line (definitely free from non-LTE effects) or the OH pure rotation lines in the IR (Sauval et al. 1984). On the other hand, equivalent widths in the integrated flux spectrum that are computed assuming LTE are consistent with that abundance (e.g. Biémont et al. 1991, Kiselman 1993, King & Boesgaard 1995). This has led most abundance analysts to keep to the LTE approximation regardless of the other evidence. It should be noted that uncertainties in atomic elling. Some workers (e.g. Tomkin et al. 1992 and Takeda 1994) have used cross sections for inelastic collisions with neutral hydrogen atoms that are sufficiently large to give results close to the LTE prediction. As will be discussed later, we suspect these cross sections to be overestimations. The "non-LTE" results referred to here are the ones which predict considerable departures from LTE, like those of Kiselman (1993).

---

[1] The abundance is defined as $A(O) = \lg N(O)/N(H) + 12$.

This paper tests the hypothesis that realistic, three-dimensional, model photospheres can reconcile the conflicting data. The reconciliation could be accomplished in several ways:

1. The triplet becomes weaker in the 3D models, while preserving its $\mu$ dependence.
2. The triplet changes its $\mu$ dependence in the 3D models so as to compromise the earlier non-LTE results.
3. The forbidden line and the molecular lines become stronger in the 3D models, indicating a lower oxygen abundance which would make the non-LTE triplet results fit observations.

We investigate the 3D non-LTE formation of the 777 nm triplet and treat the other lines in LTE.

## 2. Methods

*2.1. Granulation models*

We use granulation snapshots from simulations by Nordlund & Stein (1989, 1991, 1995). These models reproduce the observed granulation pattern quite accurately, both with respect to statistical measures such as intensity contrast and intensity power spectra, and with respect to the subjective impression of synthetic images convolved with instrumental and (terrestrial) atmospheric seeing functions.

The snapshots consist of velocity data and three thermodynamic quantities on a $125 \times 125 \times 82$ grid, corresponding to a $6 \times 6$ Mm square on the solar surface and a depth of 3.2 Mm. The original data was interpolated onto a $64 \times 64 \times 55$ grid in order to ease memory requirements by decreasing the number of points in the horizontal directions while assuring small enough spacing for the radiative transfer calculations in the vertical direction, where the upper 1 Mm (35 points) of the total simulation depth was used.

In this paper, results for two individual snapshots will be presented. These were selected to be sufficiently separated in simulation time to be independent, and they are also taken at approximate opposite phases in the oscillation that is present in the simulations (these oscillations correspond to the solar p-mode oscillations).

As a reference one-dimensional model, the semi-empirical model of Holweger & Müller (1974) (here denoted HM) was used. The model was expanded to a "3D" model with dimensions $3 \times 3$ in the horizontal plane so granulation snapshots. Velocity fields or microturbulence were introduced as described below.

Figures 8 and 9 illustrate the temperature structures of the 1D and 3D models, see also Fig. 1.

## 2.2. Solving the 3D non-LTE problem

We solve the three-dimensional non-LTE problem in the two-level-atom approximation using the correction scheme of Nordlund (1985, 1991) to find the line-source function. After convergence, the results are used to to calculate line profiles for each (x,y) point and several angles $\Omega$.

### 2.2.1. Assumptions and approximations

In the two-level-atom approximation, the line source function $S_L$ is related to the Planck function $B_\nu$, and to the line-averaged mean intensity in the line $J_L$ by (cf. Mihalas 1978, p. 336)

$$S_L = \frac{J_L + \varepsilon' B_\nu}{1 + \varepsilon'} \tag{1}$$

where $\varepsilon' = C_{UL}(1 - e^{-h\nu/kT})/A_{UL}$ is a measure of the destruction probability of line photons.

We assume complete redistribution, so $J_L$ is an average over angles ($\Omega$) and frequency:

$$J_L = \int_\Omega \int_\nu I(\nu, \Omega) \phi(\nu, \Omega) d\nu d\Omega \tag{2}$$

where $\phi(\nu, \Omega)$ is the line absorption and emission profile. ($\phi$ is dependent on $\Omega$ when there is a macroscopic velocity field present.)

The line opacity is calculated in LTE, which means that we assume that the atomic levels involved in the line transitions are populated according to the Boltzmann and Saha formulae. For the lower level in the O I 777 nm triplet case this seems to be a reasonable approximation according to the detailed modelling of Kiselman (1993). For the upper level this is obviously a less good approximation since we expect a significant non-LTE effect. However, as regards the line opacity, departures from LTE come in only as a correction to the small stimulated-emission correction.

The background source function, $S_c$, is assumed to be equal to the Planck function $B_\nu$. The background continuous opacities are calculated with the help of a package stemming from the model atmosphere code of Gustafsson et al. (1975).

Natural line broadening was included according to the life times of the atomic levels involved. Van der Waals damping was treated according to the Unsöld (1955) approximation, without any enhancement factors.

Unless otherwise noted, we have used the oxygen abundance of 8.93, that is recommended by Anders & Grevesse (1989) and largely based on OH line data and the HM lower abundance. Grevesse et al. (1994) cite a value of $8.87 \pm 0.07$, which is derived using adjusted photospheric models. We use the 8.93 as the reference value here to be consistent with the HM model and the $A_{ul}$ values used for the OH and [O I] lines.

### 2.2.2. The iteration procedure

For numerical reasons, it is convenient to work in the average of incoming and outgoing intensities (Feautrier transformation)

$$P_\Omega = \int_\nu \phi_\nu \frac{1}{2}[I(\nu, \Omega) + I(-\nu, -\Omega)]d\nu. \tag{3}$$

Then

$$J_L = \int_\Omega P_\Omega \frac{d\Omega}{4\pi}. \tag{4}$$

The problem is to find $S_L$. This is done in an iterative procedure where corrections $\delta S_L$ are computed. The basis of the method is to solve one one-dimensional two-level-atom problem along each ray ($\Omega$). This is done using the approximate lambda operator, $\Lambda_\Omega^\dagger$, of Scharmer (1981, 1984).

1. Solve the transfer equation with the current estimate of $S_L$ to get $P_\Omega^{(n)}$.
2. Compute corrections $\delta P_\Omega$ by solving the system

$$\Lambda_\Omega^\dagger \delta P_\Omega = P_\Omega^{(n)} - P_\Omega^{(n-1)} \tag{5}$$

3. The corrections to the line source function are given by

$$\delta S_L = \delta J/(1+\varepsilon') = \int_\Omega P'_\Omega \frac{d\Omega'}{4\pi}/(1+\varepsilon'). \tag{6}$$

4. Solve the transfer equation once more with the new $S_L$ to get a new $J_L$ which is used to update $S_L$. This corresponds to an ordinary $\Lambda$-iteration, and is necessary to prevent the development of small scale instabilities in the source function corrections $\delta S_L$.

Typically, we iterate until the relative corrections $\delta S_L/S_{rmL} < 10^{-3}$. After convergence, we can use the final $S_L$ for the computation of spectral lines at any $\phi$ and $\mu$ angles.

During this procedure, much interpolation is needed to transform data back and forth between different angles. It is important that this interpolation is done in an exactly reversible way. To that end we use Fourier interpolation and we also minimise the number of required interpolations by only interpolating the simulation data, $S_L$ and $\delta S_L$. We calculate line and continuous opacities, $\varepsilon'$, etc. anew for each angle.

The coding was done in the Interactive Data Language (IDL), with the exception of the calculation of the $\Lambda^\dagger$ operator that was done in an external FORTRAN routine. The use of IDL makes possible a convenient stepwise development and offers simple ways to check intermediate results and handle input/output. The disadvantage is the lower speed compared to, for example, FORTRAN code. Its successful use in heavy numerical work depends on how well the problem is possible to express as matrix operations. Some compromise had to be made in this respect, partly because of practical memory limits.

### 2.3. Computing final results

Once the line source function iterations have converged, we use the final $S_L$ estimate to formally solve the transfer equation. We use partly the same procedures for this as for the iterations, but we can now generally afford more frequency and angle points. For the results presented in this paper we have used five $\mu$ points and, to get some more statistics from each snapshot, two oppositely directed azimuthal $\phi$ directions. For LTE calculations, we enter this stage immediately with $S_L = B_\nu$.

The resulting set of line profiles, one for each (x,y) point and angle, can finally be used to calculate equivalent widths in individual points, in integrated intensity for each $\mu$, and in the integrated flux.

**Table 1.** Line parameters

| line | $\lambda, \lambda^{-1}$ (air) | $E_l$ [eV] | $g_l$ | $g_u$ | $A_{ul}$ [s$^{-1}$] |
|---|---|---|---|---|---|
| O I | 777.5 nm | 9.15 | 5 | 3 | $3.55 \cdot 10^7$ |
| O I | 777.4 nm | 9.15 | 5 | 5 | $3.55 \cdot 10^7$ |
| O I | 777.2 nm | 9.15 | 5 | 7 | $3.55 \cdot 10^7$ |
| [O I] | 630.0 nm | 0.00 | 5 | 5 | $5.95 \cdot 10^{-3}$ |
| OH | 765.7 cm$^{-1}$ | 1.20 | 44 | 48 | 195.5 |
| OH | 780.9 cm$^{-1}$ | 1.69 | 48 | 50 | 207.7 |
| OH | 919.0 cm$^{-1}$ | 2.42 | 64 | 66 | 352.2 |
| OH | 928.7 cm$^{-1}$ | 3.20 | 72 | 74 | 342.1 |

## 3. The O I 777 nm triplet

### 3.1. Modelling and input data

The lower level of these lines is of high excitation energy (9.15 eV). We use the theoretical oscillator strengths of Biémont et al. (1991) and the electron collisional rate coefficients computed by Carlsson & Judge 1993 – these data were also used by Kiselman (1993).

The expected departures from LTE are caused by the escape of 777 nm line photons, causing $S_L$ to fall below lence of 1.1 km s$^{-1}$ was employed for the HM results. Observed values for the atomic lines are from Kiselman (1993) and from Sauval et al. (1984) for the OH lines

| line | | LTE | $W_\lambda$ [pm] snap 1 | snap 2 | HM | obs |
|---|---|---|---|---|---|---|
| O I | 777.5 nm | | 6.43 | 6.16 | 6.75 | 4.94 |
| O I | 777.4 nm | | 8.07 | 7.74 | 8.51 | 6.38 |
| O I | 777.2 nm | | 9.26 | 8.88 | 9.67 | 7.24 |
| [O I] | 630.0 nm | * | 0.62 | 0.66 | 0.55 | 0.58 |
| OH | 765.7 cm$^{-1}$ | * | 23.84 | 28.46 | 23.04 | 21.40 |
| OH | 780.9 cm$^{-1}$ | * | 14.52 | 20.38 | 12.83 | 10.21 |
| OH | 919.0 cm$^{-1}$ | * | 7.06 | 7.66 | 5.67 | 4.39 |
| OH | 928.7 cm$^{-1}$ | * | 1.55 | 1.67 | 1.21 | 0.85 |

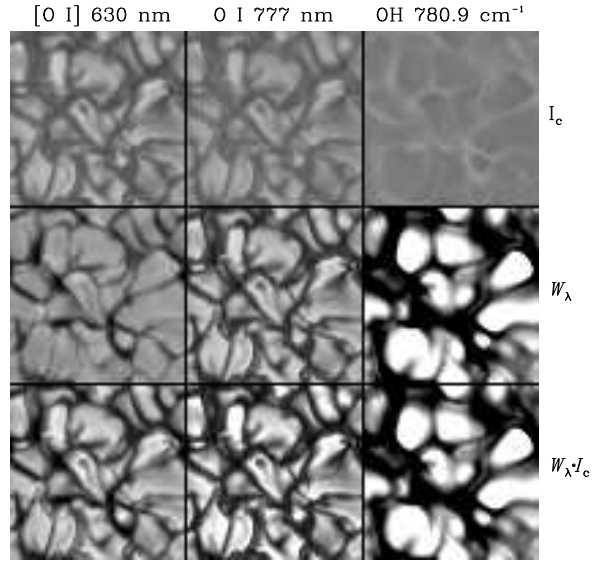

**Fig. 1.** Results for 3D snapshot number 1 and three different lines displayed as maps. Upper row: continuum intensity. Middle row: equivalent width (light means stronger line, dark is weaker line). Lower row: The intensity-weighted equivalent width. All frames have been given the same relative intensity scale to allow comparison of contrasts

its LTE value of $B_\nu(T)$. This will lead to stronger lines than in LTE. The only plausible way to have the triplet formed close to LTE is to have high enough collisional rates in the line transitions. This would require electron collisional cross sections several orders of magnitude larger than those used here or a substantial contribution from collisions with neutral atoms (hydrogen). The importance of the latter is a matter of discussion (see, e.g., Lambert (1993)). Kiselman (1993) showed that even if one accepts the (probably too large) cross section estimates introduced to non-LTE spectral modelling by Steenbock & Holweger (1984), this is not enough to totally thermalise the line

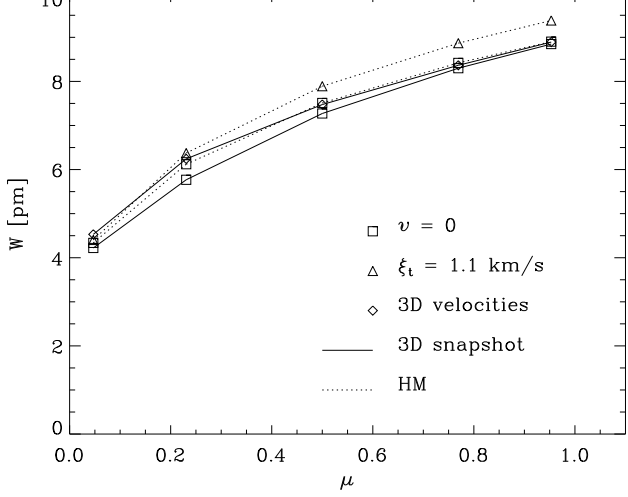

**Fig. 2.** Different treatments of the velocity field

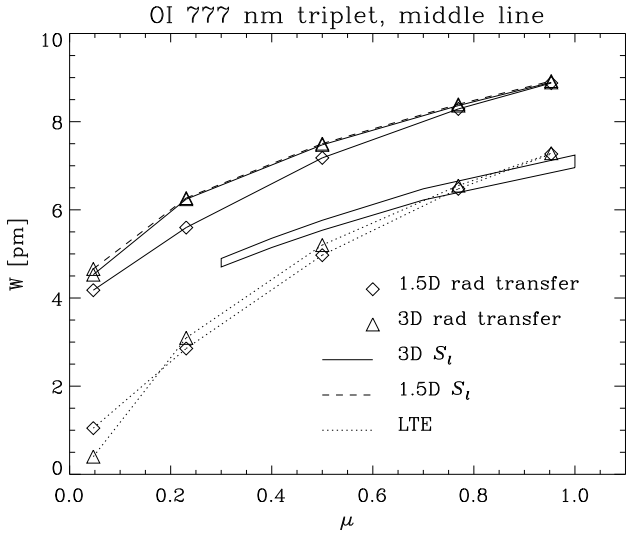

**Fig. 3.** Various treatments of radiative transfer. The band indicates observations of Müller et al. (1967) with ±2% internal error in $W_\lambda$

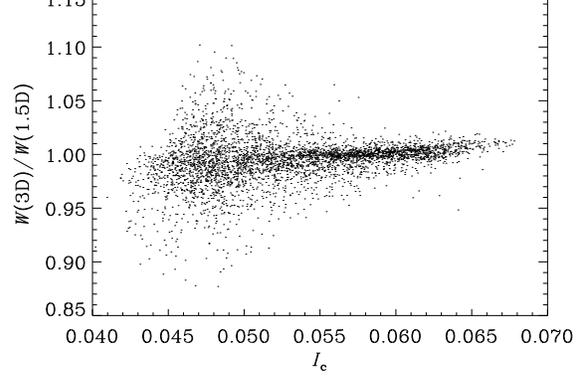

**Fig. 4.** Comparison of the equivalent widths of the middle 777 nm triplet line as calculated for snapshot 1 in 3D and in 1.5D. The ratio $W_\lambda(3D)/W_\lambda(1.5D)$ is plotted as a function of the continuum intensity (vertical rays)

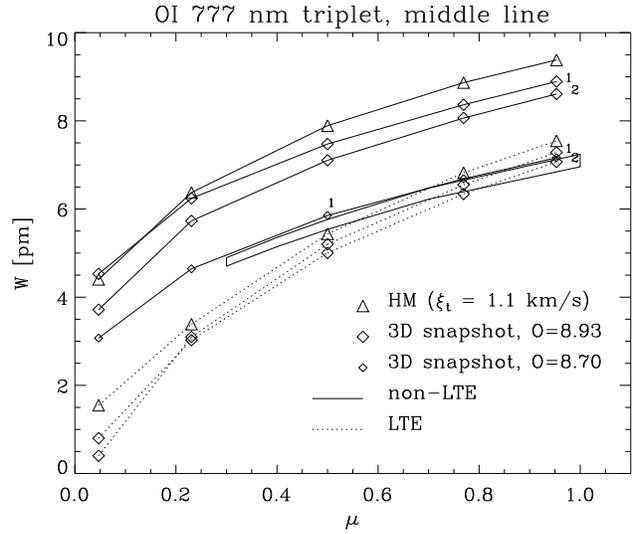

**Fig. 5.** Comparison between different 3D simulations and the Holweger-Müller 1D semi-empirical photospheric model. The band indicates observations of Müller et al. (1967) with ±2% internal error in $W_\lambda$

transitions. Tomkin et al. (1992) artificially set the hydrogen collision cross sections to five times the electron ones and thus essentially got LTE results. When the LTE approximation is used here, it should in light of this discussion be seen as a limiting case to possible line-formation circumstances and not as a physically viable alternative to non-LTE modelling.

The middle line of the 777 nm triplet was used in most of the experiments described here. The source function of this line is expected to be closest to the two-level approximation, regardless of the amount of coupling in the fine-structure of the upper level term of the triplet. We display the results in the form of plots showing the variation of the line equivalent width $W$ with position on the solar disk, $\mu$.

### 3.2. Impact of the velocity field

Fig. 2 shows the results of experiments with different treatments of the velocity field. For the 3D snapshot, the line calculation has been made with the original velocity field and with all velocities set to zero. Results for the HM model are presented with no velocity field, and with a $1.1\,\mathrm{km\,s^{-1}}$ microturbulence that is typical of the values used in abundance analyses.

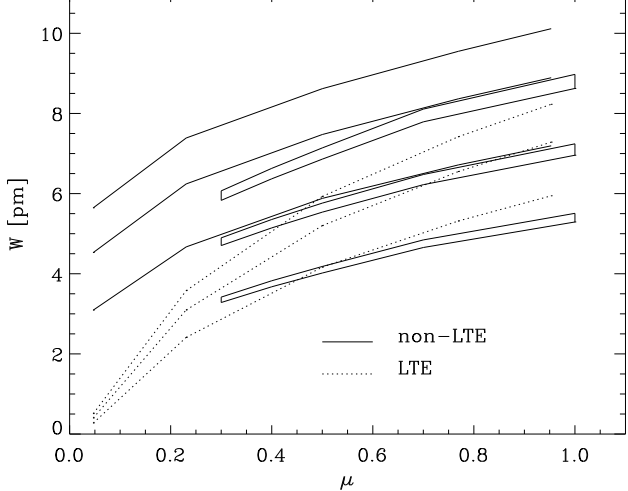

**Fig. 6.** All three lines in the 777 nm triplet: non-LTE and LTE results. The bands indicate observations of Müller et al. (1967) with ±2% internal error in $W_\lambda$

Takeda (1994) proposed that the triplet is formed close to LTE and that the problems with the centre-to-limb variations are due to the inadequacy of the classical models' microturbulence concept. The current results make this explanation unlikely since the difference in $\mu$ dependence is small between the various kinds of velocity field treatments.

The similarity between the 3D snapshot result and the HM result without microturbulence could imply that one problem with the latter is indeed the use of a microturbulence.

### 3.3. 3D effects on line formation

The linestyles in the plot of Fig. 3 signify different ways of computing the line source function, $S_L$. In LTE this is given by the Planck function. The 1.5D non-LTE method is the same as the 3D method, except that the horizontal distance between grid points is considered to be infinite. This effectively transforms the inhomogeneous 3D snapshots to an ensemble of 1D model photospheres. The plot symbols indicate how the formal radiative transfer was solved once $S_L$ was fixed.

It is obvious that the 3D non-LTE effects on $S_L$ are small in the regions where this line is formed. Figure 4 illustrates this by showing the ratio $W_\lambda(3D)/W_\lambda(1.5D)$ for vertical rays. The equivalent widths are indeed very similar in the two cases. There is a small trend for the ratio to increase when going from dark to bright regions. This is a natural consequence of the darker regions being illuminated by bright granules and the corresponding dilution of the radiation field over the latter. A locally increased $J_L$ leads to a weaker line according to Eq. 1.

and the 3D results towards smaller $\mu$ implies that three-dimensionality has some importance for the line transfer close to the limb since each ray passes through regions of the atmosphere of different character. The $\mu$ dependence resulting from the 3D treatment seems to give a better correspondence with the observations.

### 3.4. 1D vs. 3D photospheric models

Figure 5 compares results from two 3D snapshots and the semi-empirical 1D HM model. Snapshot 2 contains a fairly large dark region, making the triplet weaker than in snapshot 1. The difference between the snapshots illustrates that one or two of these is not a sufficient number for precise comparisons of equivalent widths.

The lines are somewhat weaker in the 3D snapshots than in the HM model photosphere, both in LTE and in non-LTE, but not enough to make the non-LTE curve fit the observations. This was also the result of the less ambitious calculations of Kiselman (1993). In order to approach the observational curve, the abundance must be lowered towards 8.70, for which results are also shown in the figure. Note how well this curve reproduces the observational $\mu$ dependence.

### 3.5. All three triplet lines

Figure 6 shows 3D snapshot results for all three triplet lines, both in LTE and in non-LTE, together with observations.

### 3.6. Accuracy of the 3D results

The interesting conclusion that the 1.5D approximation results in $S_L$ very close to the 3D result could make us concerned whether a sufficient number of angles were used in the iterations. To check whether two $\mu$ points are enough, a test with three such points was performed, once again for the middle of the triplet lines. The outcome of this test was reassuring, since the differences in $S_L$ amounted to at most 5% in the upper part of the simulation. The resulting change in the flux equivalent width was very small: +0.02%.

We take this as evidence that two $\mu$ points are indeed enough in this case. We hope, however, that future work can be more generous with angular points so that this cause for concern is removed.

The resampling of the simulation grid from 125 × 125 in the horizontal dimensions to 64 × 64 could also be a matter of concern. To check whether this is to sparse to sample the velocities correctly, we made tests with full resolution on subsets of the data. We found no significant differences.

## 4.1. Modelling and input data

The forbidden oxygen line at 630 nm has been extensively used for abundance analysis in cool stars. It was considered as a prime indicator for the solar oxygen abundance by Lambert (1978). It is definitely formed very close to LTE and its oscillator strength is fairly well-known. Some different $f$ values are reviewed by Kiselman (1993). We choose here to use the value of Lambert (1978), since this has been shown to be consistent with a solar abundance of 8.92-8.93 and photospheric models similar to the HM model.

## 4.2. Results

As can be seen in Table 2, the line is about 10 % weaker in the 3D snapshots than in the HM model. Taken by itself this would correspond to a lowering of the solar oxygen abundance with about 0.08 dex. Anyway, the results indicate that this line may not be insensitive to the details of granulation.

## 5. OH lines

### 5.1. Modelling and input data

We have studied the formation of four of the pure rotational OH lines that were observed and analysed by Sauval et al. (1984). The lines were chosen to span a range in strength and excitation energy.

The line calculations were made in LTE. The possibility of LTE departures for molecular lines is a largely uninvestigated one. We note that Hinkle & Lambert (1975) argue that rotational and vibration levels in a molecule like OH are probably collisionally populated, which would mean that the lines are formed close to LTE. The possibility of non-LTE effects in the molecular equilibrium is another matter.

The molecular populations were computed by the continuous opacity package. Spectral line data was taken from the Sauval et al. (1984) paper.

### 5.2. Results

As seen in Table 2, our computed equivalent widths for the HM model differ from those observed by Sauval et al. (1984). Ideally, this should not be the case since we have used the same atmospheric model and an oxygen abundance close to what these authors derive. This kind of situation is not uncommon, and we cannot determine whether it is due to differences in continuous opacities, molecular equilibrium constants, different treatments of radiative transfer, or something else. We do not, however, consider these differences significant for our qualitative discussion here. A more important reason for caution is most part of the simulation snapshot, where these may be least accurate, and where the strongest lines might become optically thick at the upper boundary.

As is obvious from Fig. 1, these lines and the surrounding continuum are formed at significantly higher levels in the photosphere than the atomic lines in the visual and near IR spectral regions. At these levels, the granulation pattern is reversed: the gas temperature is lower above rising granules than above intergranular lanes. Since the formation of molecules is very sensitive to temperature, and is strongly enhanced in cool regions, the OH lines behave quite differently from the atomic oxygen lines as well as all other atomic lines observed by Kiselman (1994) in the sense that they get stronger in cool regions. Hence, the equivalent width maps in the middle row of Fig. 1 have roughly the same topology though the continuum maps in the upper row look different.

Comparing our HM and 3D snapshot results, it is interesting to note that the 1D model gives weaker lines than the snapshots in all cases. Sauval et al. demonstrated that their observations and molecular data together with the HM photosphere give consistent abundances for an impressive number of OH lines. But we *know* that the HM photosphere is a one-dimensional approximation of a three-dimensional and inhomogeneous reality, so this consistency *may* be spurious. It will, however, take more and better evidence than what is presented here to really challenge the 1D molecular results. But it does not seem unlikely that the oxygen abundance around 8.9 derived from 1D analysis could be about 0.1 dex too high.

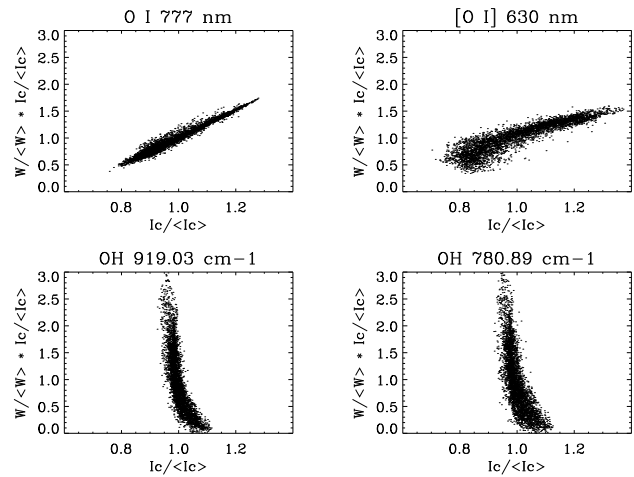

**Fig. 7.** The contribution to the line equivalent width in integrated light (vertical rays) from each $(x, y)$ point of snapshot 1, $W(x, y) \cdot I_c(x, y)$, plotted as a function of the continuum intensity, $I_c$. Intensities and equivalent widths have been normalised with their mean values and all plots have the same scaling to facilitate comparisons

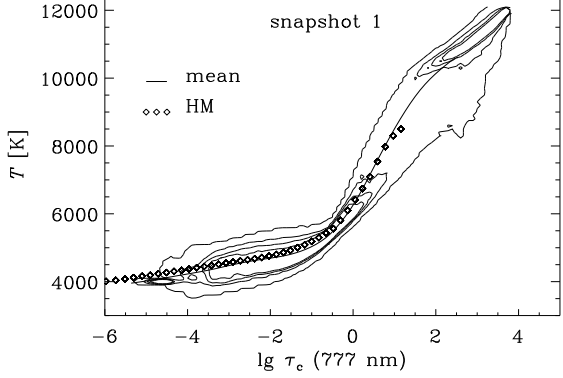

**Fig. 8.** Temperature as a function of continuum optical depth (at 777 nm) for snapshot 1 compared with the HM model and the "mean" 1D model constructed from the snapshot data.

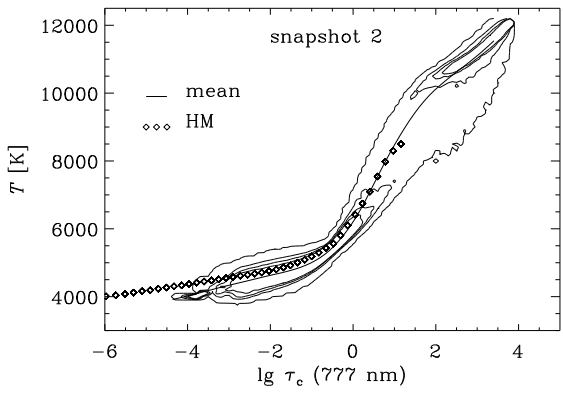

**Fig. 9.** Same as Fig. 8, but for simulation snapshot 2

## 6. More on 1D models

In order to further illustrate the difference between 1D and 3D models we constructed 1D models from the two 3D snapshots in the following way. For each 3D model, the temperature, pressure and the logarithm of the density were averaged over the continuum optical depth for three of the lines (the middle 777 nm line, the [O I] 630 nm line and the 919 cm$^{-1}$ OH line). In the forming of the averages, the contribution from each (x,y) point was weighted with the emerging continuum intensity (vertical rays). The three resulting models were similar in structure, and we also note that the weighting procedure did not make too much difference.

Figures 8 and 9 illustrate the structure of the different 3D snapshots and 1D models.

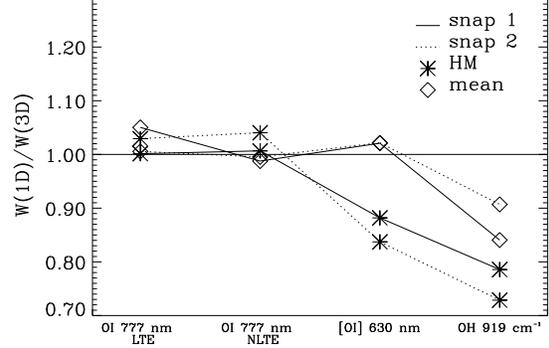

**Fig. 10.** Equivalent width ratios (vertical rays) comparing 1D and 3D models without velocity fields. Symbols indicate the 1D model and line-styles the 3D snapshot used. The "mean models" are constructed from the 3D snapshots (see text). HM is the semi-empirical model of Holweger & Müller (1974)

The line calculations in this comparison were made with zero velocity field so the differences are due only to the model structure. A comparison of the equivalent widths ($\mu = 1$) for the three lines and the different photospheric models is shown as Fig. 10. It is apparent that the molecular line is stronger in the 3D snapshots because of the cool pockets present there.

Once more, we see that snapshot 1 produces very similar triplet line strengths to the HM model when all velocity fields are set to zero. The [O I] 630 nm line comes out significantly weaker in the HM model than in the 3D snapshots or the "mean" 1D models. Note, however, that a 10 % change in equivalent width corresponds to about 20 % in abundance for the 777 nm triplet line, about 7 % for the 919 cm$^{-1}$ OH line, while the strength of the 630 nm line increases linearly with abundance.

If a microturbulence were to be introduced in the 1D models and simulation velocity fields in the 3D models, all points (but especially the 777 nm results) would move upwards in the diagram (cf. Fig. 2).

## 7. Discussion

Have the 3D models been able to reconcile theory and observations for the oxygen lines? In the introduction we listed three possible ways for this to happen. Possibility number 2 in our list has apparently been excluded – the $\mu$ dependence of the triplet lines does not differ much between the different photospheric models. We take this as an argument for that the predicted line-source-function depression via photon escape really is a proper description of the formation of the triplet lines, and that this process really occurs regardless of the atmospheric structure.

The other possibilities seem, on the other hand, to have been fulfilled – the triplet lines are systematically weaker bidden and the molecular lines are stronger. The effects are, however, not strong enough to take us the whole way and they are not totally convincing in light of the various modelling uncertainties. A larger sample of realistic 3D models is needed before the predictions of integrated light line strengths can be considered as firm.

It is interesting to note that snapshot 1 is very similar to the HM model (when no velocity fields are included) for the triplet line formation, but not for the other line. This is indeed an example of a 1D photospheric model being representative for the average properties of a 3D model in one aspect but not for others.

Figure 7 highlights the different behaviour of the lines in the 3D models (snapshot 1) by showing the intensity-weighted contributions to the integrated equivalent width as functions of continuum intensity. The colder parts of the upper photosphere contribute immensely more to the equivalent widths of the OH lines than the warmer regions. 1D models are averages and in general one could not expect them to properly account for the non-linear formation of molecules. That 1D models give weaker lines than 3D models due to this is evident in Fig. 10.

We note also that the measurement of equivalent widths is a partly subjective procedure. Comparison of the solar equivalent widths measured and cited by King & Boesgaard (1995) and Kiselman (1991, 1993) shows that they differ within about ±7 % between different authors and data sets for the triplet lines and even more for the weak 630 nm line. (The figure for the triplet lines corresponds to about ±0.06 dex in derived abundance.) There may also be systematic differences between equivalent widths measured from observed spectra and those that are computed. In the latter, all contribution from the extended line wings can be exactly accounted for, while the treatment of line wings and continuum placement is a problem in the measurement. In this context we should note that the spatially-resolved observations of the 777 nm triplet showed in the plots are from Müller et al. (1967). They are very similar to the Altrock (1968) results. King & Boesgaard (1995) essentially confirmed these results with new observations, but problems with scattered light in their spectrograph prevented improvements.

King & Boesgaard (1995) argue that stellar forbidden-line abundances may be ridden with errors as big as for those derived from the 777 nm triplet. If this is the case, and the difficulty in making data from different datasets match by changing atomic and atmospheric parameters (Kiselman 1993, King & Boesgaard 1995) makes it probable, the question of the precise oxygen abundances of solar-type stars is still unresolved.

## 8. Conclusions

In agreement with the conclusions of Kiselman (1993), we have found the shape of the variation of the O I 777 nm the temperature structure and the velocity fields of solar photospheric models. The limb-darkening function is thus mostly sensitive to the atomic quantities that govern the destruction probability of line photons. We consider the success of the non-LTE modelling in reproducing the observed dependence of line strength on $\mu$ to be compelling evidence that we have got the atomic physics right, and that these lines are not formed close to LTE.

We have not been able to reconcile the lines studied with one single oxygen abundance, even though the 3D results move things in the right direction compared to the 1D HM model photosphere. The 777 nm triplet lines do become weaker in the 3D models, and the 3D results for the O I 630 nm line and the OH lines indicate that the solar oxygen abundance could very well be less than 8.80. The remaining discrepancy between observations and theory for the 777 nm triplet could then fall within the range of observational error and errors in $f$ values. There may also be room for slightly larger collisional cross sections in the triplet line transitions to weaken the lines while saving their centre-to-limb behaviour.

The line source function of the O I 777 nm triplet is well described by the 1.5D approximation – 3D non-LTE effects do not seem to be important. This means that it is possible to solve the non-LTE problem in 1.5D in this case, permitting a much more complicated model atom. The formal radiative transfer calculation when the source function is known should then preferably be done in 3D. This result should hopefully hold also for other lines that are formed in about the same circumstances as the O I 777 nm triplet.

*Acknowledgements.* Kjell Eriksson gave helpful advice on continuous opacities. This work was supported in part by the Danish National Research Foundation through its establishment of the Theoretical Astrophysics Center.